\documentclass[fleqn,usenatbib]{rasti}

\usepackage{newtxtext,newtxmath}

\usepackage[T1]{fontenc}
\usepackage{subcaption}

\DeclareRobustCommand{\VAN}[3]{#2}
\let\VANthebibliography\thebibliography
\def\thebibliography{\DeclareRobustCommand{\VAN}[3]{##3}\VANthebibliography}

\usepackage{graphicx}	
\usepackage{amsmath}	
\usepackage{fontawesome5}
\usepackage{xcolor}

\title[Spectral Sensitivity Analysis with \texttt{PICASO}]{Exoplanet Atmosphere Spectral Sensitivity Analysis with \texttt{PICASO}: Jacobians, Linear-Gaussian Approximations, and Information Content}

\author[Batalha \& Wogan]{
Natasha E. Batalha,$^{1}$\thanks{E-mail: natasha.e.batalha@nasa.gov (NEB)}
Nicholas F. Wogan,$^{1}$
\\
$^{1}$NASA Ames Research Center, Moffett Field, CA 94035, USA
}

\date{Accepted 2026 September 08. Received 2026 September 04; in original form 2026 May 12.}

\pubyear{\the\year{}}

\begin{document}
\label{firstpage}
\pagerange{\pageref{firstpage}--\pageref{lastpage}}
\maketitle

\begin{abstract}
Designing observatory and instrument architectures for complex missions like the Habitable Worlds Observatory (HWO) requires a rapid and rigorous mathematical framework to evaluate trade-offs across bandpass, spectral resolution, and signal-to-noise regimes. While full Bayesian atmospheric retrievals are the gold standard for atmospheric inference, they are too computationally expensive to explore this full parameter space.  Here, we present a  sensitivity analysis and retrieval diagnostic toolkit within the open-source code \texttt{PICASO}. The core addition to the code base is a flexible solution for computing finite-difference Jacobian matrices, which provides a local linearization of the forward model. These Jacobian matrices characterize the sensitivity of the spectrum (either reflected, thermal, or transmission) to perturbations in any atmospheric state parameter, such as molecular abundances. Given the Jacobian for a certain reference atmosphere, our toolkit provides a suite of analytic diagnostic quantities under Gaussian assumptions for the prior, likelihood, and posterior. These include Fisher-matrix-based estimates of 1$\sigma$ constraint intervals of atmospheric parameters, averaging kernels, singular-value-decomposition diagnostics for parameter degeneracies, and an information-content metric.  
We validate the results of our toolkit by comparing them to that of a full Bayesian retrieval of  an Earth-like planet with an HWO-like facility. We show that our metrics can provide rapid understanding of model parameter sensitivity and that local linear-Gaussian approximations can diagnose dominant degeneracies seen in retrievals as well as provide first order estimates of posterior probability distributions. 
Ultimately, this provides the community with a complementary toolkit for   optimizing instrument design and conducting spectral sensitivity diagnostics.
\end{abstract}




\section{Introduction}

Determining a mission and instrument architecture that is capable of observing substellar atmosphere spectra requires an in-depth understanding of where spectral information exists. For any given substellar atmosphere, whether it be a temperate rocky planet or a large gaseous hot Jupiter, fundamental information is needed: 1) what molecular or atomic species are visible, 2) what wavelengths they appear at, 3) what spectral resolution is needed to resolve their features, and 4) what other factors such as planet or stellar properties, climate, or atmospheric scattering properties, continuum absorption, also influence the spectra. 
These types of discussions have been addressed in many technical documents outlining mission concepts with respect to their overall scientific mission objectives (e.g., for the James Webb Space Telescope (JWST) pre-launch  \citep[e.g.,][]{beichman2012science,2014PASP..126.1134B} or for the 2020 Astrophysics Decadal Survey concepts (e.g., HabEx \citep{gaudi2020habitable} and LUVOIR \citep{luvoir2019luvoir}). An in depth understanding of this information allows for scientists and engineers to conduct numerous trades between instrument properties (e.g., resolving power, signal-to-noise ratio, bandpasses) to ultimately arrive at an architecture which meets the mission's overall science objectives. 

There are two traditionally used methods for conducting this work. Historically, the most common technique utilized across planetary science, earth science, and astrophysics mission design was basic information content (IC) theory \citep{shannon1948mathematical,2000imas.book.....R,2012ApJ...749...93L}. We emphasize here that throughout the manuscript we refer to ``IC theory'' as a term to encapsulate the broad suite of diagnostics that are born from creating a linearization of a forward model and using Gaussian approximations to the prior, likelihood, and posterior. In modern statistical language, the method is closely analogous to Fisher-matrix forecasting or simply a Gaussian approximation to the posterior information gain. We appreciate that using IC theory has the potential to be confused with the scalar bit-quantity referred to as ``information content''. However, due to the popularity of the broad ``information content'' term in some remote-sensing and exoplanet communities, we continue that adoption here. We make the distinction to ``scalar IC'', when needed. Broadly speaking though, we use IC theory to differentiate itself from ``retrieval analyses''.  

More recently, with gains in computational power Bayesian-based spectral inference methods, or ``retrievals'', have risen in popularity with regards to substellar atmosphere spectral information assessment \citep[e.g.,][]{2012ApJ...753..100B, 2014ApJ...783...70L,2018AJ....155..200F}. For reasons we describe in the following subsection, these two methods provide complementary information that should be used in tandem to give mission architects an optimal toolkit to understanding spectral information. However, the availability of tools and resources for retrieval analysis \citep[see full table in][]{2023RNAAS...7...54M} has vastly out-paced the resources available for basic information content theory, as it pertains to the characterization of substellar atmospheres. In principle the same tools that can be used to compute substellar atmosphere spectra and run a retrieval can be used to conduct basic IC theory (which we again emphasize points to more broadly other local linear-Gaussian-approximation spectral sensitivity diagnostics). Yet, there is ultimately a lack of direct pipelines available to conduct these analysis. 

\texttt{PICASO} is a widely-used open source radiative transfer tool that has applicability across solar system, exoplanets, and brown dwarfs. It was first developed to compute the reflected light spectra of directly imaged exoplanets \citep{2019ApJ...878...70B} but has since expanded to many other applications including: 1) 1D radiative-convective thermochemical equilibrium climate modeling \citep{2023ApJ...942...71M}, 2) transmission spectroscopy \citep[e.g.,][]{2023AJ....165...14B,2023Natur.614..664A}, 3) thermal emission spectroscopy of exoplanets \citep[e.g.,][]{2024ApJ...973L..41I} and brown dwarfs \citep[e.g.,][]{2024AJ....167..237L}, 4) reflected light phase curves \citep{2024ApJ...976..181H}, 5) thermal phase curves \citep{2022ApJ...930...93R}, 6) retrievals \citep[e.g.,][]{2026AJ....171..227L}, and most recently 7) 1D radiative-convective disequilibrium and cloudy climate modeling  \citep{2026ApJ..1000...98M}. Although there have been individual publications that have utilized \texttt{PICASO} to conduct information content analysis \citep[e.g.,][]{2019ApJ...878...70B, 2021SPIE11823E..08B} the functionality has never been formally introduced into the code-base. 

Herein lies the driving purpose of this manuscript, which is to pair a well-vetted radiative transfer tool applicable to many areas of substellar atmosphere characterization directly with a suite of tools related to these spectral sensitivity analyses. The novelty of the work is to use traditional metrics but tailor them to the needs of the exoplanet community and wrap everything into an open-source easy-to-use framework. Some of our diagnostics, such as using singular value decomposition to assess degeneracies, have been less commonly used in the exoplanet community. We first (\S \ref{sec:overview}) provide a brief conceptual overview of IC theory and how it relates to retrieval techniques to ensure users can choose an appropriate methodology for their problem. Then, we take a deeper look at the methodology of IC theory and how it is applied within  \texttt{PICASO} \S \ref{sec:methods}. \S \ref{sec:methods} also describes all of the spectral diagnostic metrics the code provides, as well as context to ensure users know how to interpret them. Next, we apply these new capabilities specifically to the Habitable World Observatory's science driving case: observing an Earth-like planet in reflected light \S \ref{sec:feng}. We end by discussing the ultimate utility of the tool to HWO and beyond in \S \ref{sec:d_and_c}.

\section{Information Content Theory vs. Retrievals: A Conceptual Overview}
\label{sec:overview}

The foundation of classical information theory is based on Shannon entropy \citep{shannon1948mathematical,2000imas.book.....R}. It provides a mathematical representation of the initial entropy of a system, and quantifies how much of this entropy can be removed when new information is gained. Intuitively, higher entropy corresponds to broader probability distributions over possible atmospheric states and therefore greater uncertainty in the planet’s atmospheric properties. 
For example, prior to observing the atmosphere of a rocky world with HWO the entropy depends only on the prior knowledge of the system. The prior knowledge could be constraints on planet radius, the planet's orbit, and a rough estimate of the planet's temperature, given its stellar insolation flux. The prior knowledge on other parameters of interest, however, such as the dominant absorbing species or the abundance of trace species is not known. In this case, we have wide uninformative priors. For example, typical abundance priors for exoplanet retrievals can span up to 10 or more orders of magnitude \citep[e.g.,][]{2013ApJ...778..183L}. An example of a regime where we have informative priors would be a remote sensing observation of CO$_2$ abundance on Earth. The prior used for a retrieval of CO$_2$ abundance might only be a few percent to capture temporal or spatial variations  \citep[e.g.,][]{saigusa2008temporal}. In either case, once an observation is made with corresponding measurement uncertainty and wavelength coverage, the overall entropy can be reevaluated. The final scalar information content (scalar IC), measured in bits, is the Gaussian entropy reduction and represents difference between the initial and final entropy. It is important to note that scalar IC, like atmospheric retrievals, is always evaluated relative to prior knowledge so input priors can dominate outcomes. 

This methodology has been leveraged across many realms of science. For example, 
\begin{itemize}
    \item \citet{2010JQSRT.111.1296K} used it to determine the observing bandpasses for \textit{Orbiting Carbon Observatory} to measure CO$_2$ on Earth
    \item \citet{2012ApJ...749...93L} used it to determine the spectral information encoded in some of the first exoplanet transmission spectroscopy models
    \item \citet{2017AJ....153..151B}, \citet{2017ApJ...835...96H} and \citet{2018ApJ...856L..34B} (among others) used it to evaluate optimal JWST observing strategies for characterizing exoplanet atmospheres
    \item \citet{2024AJ....167..240F} used it to determine the importance of optical transmission spectrum data taken by the Hubble Space Telescope
    \item \citet{2021SPIE11823E..08B} used it to aid in the design of  SCALES/Keck, a new ground-based instruments for substellar atmosphere characterization 
\end{itemize} 

Overall, it is a well-vetted methodology that has proved its utility throughout many decades. Nevertheless, IC theory has fundamental limitations. There are primarily two assumptions that necessitate the use of a complementary method, especially for the case of prior-dominated observing scenarios, such as the HWO case described above. The assumptions are: 1) that the model can be linearized at the atmospheric reference state under consideration, and 2) that the priors, likelihood, and posterior probability distributions are Gaussian. 
Both of these conditions are only sometimes valid in  cases of substellar atmosphere characterization, especially as it relates to the characterization of temperate rocky worlds with the Habitable Worlds Observatory. This is in part why Bayesian-based spectral inference, or retrieval methods, have risen in popularity. 

Bayesian-based spectral inference has routes that are similar to that of information theory. Both are grounded in Baye's Theorem and on defining a set of parameters and associated atmospheric model. And, both depend on defining a set of priors for each of these atmospheric parameters. 
There are several in depth reviews of retrievals as they relate to substellar atmosphere characterization \citep[e.g.,][]{2019ARA&A..57..617M,2020SSRv..216...82B,2023RNAAS...7...54M}. Therefore, here we limit the discussion to how retrievals overcome the fundamental limitations of traditional IC theory. The ultimate result of a retrieval is the posterior probability distribution, the probability the model parameters exist given the data \citep{gregory2005}. There are many different kinds of retrieval sampling algorithms (e.g., MCMC \citep{chib1995understanding, neal2011mcmc,hoffman2014no} and Nested Sampling \citep{skilling2004nested, skilling2006nested, sivia2006data, feroz2008, feroz2019}). At their core they all have the ability to 1) adequately explore a high-dimensional and oftentimes large prior volume space, and 2) approximate a posterior probability distribution for each model parameter of interest. Because they can explore a full range of prior volume space, they can (with varying levels of success) capture model non-linearity. Additionally, all algorithms have some ability to capture non-Gaussian posterior probability distributions. Types of non-Gaussian distributions common to observations of substellar atmospheres include: 1) a bi-modal distribution, 
2) an upper/lower limit distribution, where posterior probability volume is reduced from the prior but does not represent a fully constrained solution, and 3) a degenerate or co-varying distribution, where the posterior probability is driven by one or more other parameters in your system. These complex posterior probability shapes cannot be captured with traditional IC theory. Several cases related to the retrievals of substellar atmosphere characterization have some degree of non-gaussian posterior probability distribution within the set of atmosphere model parameters \citep[e.g.,][]{2019AJ....157..206W}. As such, a natural question to consider is: \textit{what then, is the utility of IC theory? }

The main motivation to use a combination of both techniques is the computation cost associated with retrievals. One retrieval run often requires hundreds of thousands of model evaluations to sufficiently cover the vast prior probability space \citep{2019ARA&A..57..617M}. Therefore, the calculation of one posterior probability distribution for a single planet case and one set of resolution, wavelength, and signal-to-noise instances could take dozens or even hundreds of computation hours. For the methodology we describe here, the maximum number of model evaluations that IC theory requires is $=2\times$ the number of atmospheric model parameters. After these initial evaluations are completed, each combination of resolution, wavelength, and observation error instance  runs ``instantaneously'' (i.e., as fast as a simple set of matrix arithmetic can be done for a modern computer). Therefore, one strategy would be to use IC theory metrics to cover broad parameter space in atmospheric reference state and instrument parameters. Spanning a range of atmospheric reference states is necessary to guard against assumptions of model linearity. Then, spanning a range in instrument parameters  would allow one to identify information cliffs at these reference states that would point to important cases to consider. Finally, retrieval runs could be strategically run on isolated cases to report final expectations for probability distributions.  

Ultimately, we emphasize that IC theory does not replace retrieval studies. Retrieval studies are needed to ground IC theory in the reality of what posterior probability distributions can be realized for a given set of parameters. This is why in this new software update, we focus on the utility of these spectral sensitivity diagnostics in its ability to: 
\begin{enumerate}
    \item Determine where in wavelength space, fundamental spectral sensitivity exists for a given set of model parameters 
    \item Identify basic geometric degeneracies between model parameters 
    \item Quantify relative information gain and loss across large regions of parameter space (wavelength, resolution, observational error)
\end{enumerate}

\section{Methods: Shannon Information Content Theory}
\label{sec:methods}
There are many comprehensive sources on information content theory specifically relating to atmospheric inference problems \citep[e.g.,][]{2000imas.book.....R}. Here for completeness we highlight the most relevant mathematical equations, forgoing a complete overview of the intermediate derivations required to arrive at them. The basic derivations adopted here are from the more comprehensive derivations in \citet[Section 2.3.2 ``The Bayesian approach to inverse problems''][]{2000imas.book.....R}. In the following subsections we describe the implementation of these calculations within the \texttt{PICASO} framework. 

The starting components of Shannon information content theory are: 1) a set of assumed model parameters, 2) a model which, given this set of parameters, describes the predicted observable (i.e., spectrum),  3) an initial guess of the true atmospheric state, and 4) the observations. We define these and provide examples for each:
\begin{enumerate}
    \item $\mathbf{x}$: a vector of $n_p$ model parameters. Examples include atmospheric abundances, a parameterization for the temperature-pressure profile, and cloud parameters. In defining this vector we assume that it ``perfectly'' describes any true atmospheric state, which is of course a simplification. 
    \item $\mathbf{F(x)}$: the model, which here is \texttt{PICASO}'s calculation of a spectrum (either in reflection, transmission, or thermal emission). 
    \item $\mathbf{x_a}$: the atmospheric reference state. We strive to linearize the model ($\mathbf{F(x)}$) at this reference state, making it important that $\mathbf{x_a}$ be close to the true state. If one were conducting a trade study using a full retrieval analysis, $\mathbf{x_a}$ would be analogous to the spectrum that is chosen as the simulated observation. 
    \item $\mathbf{y}$: the spectral observation and associated error bars for a given number of wavelength channels ($n_\lambda$). For trade studies the observation is often simulated using tools such as \texttt{PandExo} \citep{2017PASP..129f4501B}. 
\end{enumerate}
Using a first-order Taylor expansion around the reference state $\mathbf{x_a}$, we can approximate the observation as 
\begin{equation}
    \mathbf{y} = \mathbf{F(x)} \approx \mathbf{F(x_a)} + \mathbf{K(x - x_a)}
    \label{eqn:linear}
\end{equation}
where $\mathbf{K}$ is the ``Jacobian'' matrix of size $n_\lambda \times n_p$. Note from this point forward, $\mathbf{K_a}$ is used to emphasize the evaluation of the Jacobian at the reference state $\mathbf{x_a}$. The Jacobian matrix is at the foundation of information content theory and describes the sensitivity of the model to each of our chosen parameters: 
\begin{equation}
    K_{ij} = \frac{\delta F_i(\mathbf{x})}{\delta x_j}
\end{equation}
where $j$ represents each of the $n_p$ model parameters and $i$, each of the $n_\lambda$ wavelength channels. In \S \ref{sec:jacobian} we elaborate on the calculation of the matrix, and detail how it is computed in \texttt{PICASO}. For continuity here, we follow along to the ultimate derivation of information content. 

Information content ($H$), measured in bits, describes how the observation increases the state of knowledge, relative to the prior. As previously stated, it is the Gaussian entropy reduction and is calculated by comparing the entropy of probabilities that a given atmospheric state exists before and after such observation: 
\begin{equation}
    H = \mathrm{entropy [P(}\mathbf{x}\mathrm{)]} - \mathrm{entropy [P(}\mathbf{x|y}\mathrm{)]}
    \label{eqn:entropy}
\end{equation}
where
\begin{align}
      \mathrm{P(}\mathbf{x}\mathrm{)} &\propto \exp \left[ -0.5(\mathbf{x}-\mathbf{x_a})^\mathrm{T} \mathbf{S_a}^{-1} (\mathbf{x}-\mathbf{x_a})
      \right]  \label{eqn:probs1} \\
      \mathrm{P(}\mathbf{x|y}\mathrm{)} &\propto \exp \left[ -0.5 \mathbf{J}(\mathbf{x}) \right].  \label{eqn:probs2}
\end{align}

In the expression of $\mathrm{P(}\mathbf{x|y}\mathrm{)}$, $\mathbf{J}(\mathbf{x})$ is the cost function, which can be defined as: 

\begin{equation}
    \mathbf{J}(\mathbf{x}) = (\mathbf{y} - \mathbf{F(x)})^\mathrm{T} \mathbf{S_e}^{-1}(\mathbf{y} - \mathbf{F(x)}) + (\mathbf{x} - \mathbf{x_a})^\mathrm{T} \mathbf{S_a}^{-1}(\mathbf{x} - \mathbf{x_a})
    \label{eqn:cost}
\end{equation}
where by substituting in our linear assumption $\mathbf{F(x)} \approx \mathbf{F(x_a)}+\mathbf{K_a(x-x_a)}$, we can arrive at an analytical solution. Ultimately, the cost function balances the two sources of information available to constrain the atmospheric state. Note that traditional Optimal Estimation algorithms \citep{shannon1948mathematical} for spectral inference rely on minimizing this cost function and have been widely used in high signal-to-noise,  high resolution regimes where model linearity and Gaussianity can often be assumed. The first term on the left-hand side is effectively the ``measurement term'' and only considers the observation ($\mathbf{y}$), measurement error ($\mathbf{S_e}$), the model ($\mathbf{F(x_a)}$ and $\mathbf{K_a}$), and the atmospheric state vector ($\mathbf{x}$). The second term on the right-hand side is effectively the prior and only considers the how far away the atmospheric state vector ($\mathbf{x}$) has deviated from what was known before the measurement (the initial state vector, $\mathbf{x_a}$) as well as the prior confidence on each parameter ($\mathbf{S_a}$). The square root of the diagonal elements of $\mathbf{S_a}$ are the Gaussian prior 1$\sigma$ widths that are centered at $\mathbf{x_a}$ for each of the atmospheric parameters of interest. Plugging Eqn. \eqref{eqn:probs1} and \eqref{eqn:probs2}  into Eqn. \ref{eqn:entropy}, we can now arrive at an expression for $H$ by assuming Gaussian probability distributions. Ultimately, $H$ is written as: 
\begin{equation}
\refstepcounter{equation}
    H = \frac{1}{2} \ln  \left| \mathbf{\hat{S}}^{-1} \mathbf{S_a} \right| 
    \tag*{\small\href{https://github.com/natashabatalha/picaso/blob/16b428fb299bcf6663a29b828c0fdef376c6af66/picaso/information_content.py\#L436}{\color{gray}\faCode}\quad(\theequation)} 
     \label{eqn:H}
\end{equation}
where we have introduced a new term, $\mathbf{\hat{S}}$, the posterior covariance matrix. The posterior covariance matrix describes the 1$\sigma$ uncertainties and correlations of the atmospheric state vector after the measurement is made. It is expressed as: 
\begin{equation}
\refstepcounter{equation}
    \mathbf{\hat{S}} = \left( \mathbf{K_a}^\mathrm{T} \mathbf{S_e}^{-1} \mathbf{K_a} + \mathbf{S_a}^{-1} \right)^{-1}.
    \tag*{\small\href{https://github.com/natashabatalha/picaso/blob/68915f96148b32ee7ade1f343a43a7c77427b2b8/picaso/information_content.py\#L414}{\color{gray}\faCode}\quad(\theequation)}
    \label{eqn:s_hat}
\end{equation}
$\mathbf{\hat{S}}$ is a parameter of interest in information theory as it can be directly compared to the 1$\sigma$ constraint intervals that are the outcome of a full Bayesian retrieval analysis. In more modern statistics $\mathbf{\hat{S}}$ is often referred to as the inverse of a prior-regularized Fisher information matrix (FIM). The FIM, being the measurement term on the left hand side ($\mathbf{K_a}^\mathrm{T} \mathbf{S_e}^{-1} \mathbf{K_a}$). Similarly, the resultant 1$\sigma$ constraint intervals are often described as the marginal standard deviations obtained from the diagonal elements of the Gaussian posterior covariance matrix.  

Similar to the cost function, the structure of the equation is intuitive: the left representing the measurement term and the right representing the prior knowledge term. The goal is to minimize the posterior covariance. First, consider the measurement term (the FIM),  $\mathbf{K_a}^\mathrm{T} \mathbf{S_e}^{-1} \mathbf{K_a}$, which describes the ``strength of the observation''. The wavelength regions where parameter sensitivity, described by the Jacobian ($\mathbf{K_a}$), is the highest and the measurement error ($\mathbf{S_e}$) is the lowest, will  drive the resultant covariance to lower values. Now, consider the addition of the prior matrix, $ \mathbf{S_a}$. When our initial knowledge is vague (i.e., $\mathbf{S_a}$ is high) and the strength of the observation is high, then $ \mathbf{S_a}^{-1} << \mathbf{K_a}^\mathrm{T} \mathbf{S_e}^{-1} \mathbf{K_a}$ and our posterior covariance matrix becomes ``data-driven''. Conversely, when a solution is ``prior-dominated'', the data cannot overcome our initial assumptions. Either the measurement error is too high in regions where model sensitivity is low, or simply we do not have parameter sensitivity in the wavelengths we are conducting an observation. 

The last Shannon information content parameter that we emphasize is the averaging kernel matrix, $\mathbf{A}$. It provides a singular metric to assess the data versus prior influence. $\mathbf{A}$ is computed as 
\begin{equation}
\refstepcounter{equation}
    \mathbf{A} = \left( \mathbf{\hat{S}} \mathbf{K_a}^\mathrm{T} \mathbf{S_e}^{-1} \right) \mathbf{K_a}.
    \tag*{\small\href{https://github.com/natashabatalha/picaso/blob/16b428fb299bcf6663a29b828c0fdef376c6af66/picaso/information_content.py\#L428}{\color{gray}\faCode}\quad(\theequation)}
\end{equation}
$\mathbf{A}$ is a $n_p \times n_p$ sized matrix where the diagonal elements are now simply valued 0--1. Values approaching 1 represent fully-data driven parameters. Values approaching 0 represent fully prior-driven parameters. A powerful metric then becomes the sum of the diagonal elements of $\mathbf{A}$, which give us the total degrees of freedom that our observation is sensitive to. 

Ultimately, Shannon information content theory offers a set of simple-to-compute metrics that can be used to pre-diagnose or intuitively understand the outcome of a full Bayesian retrieval. Lastly, we  emphasize the utility of the metrics described here for comparing outcomes of: 1) different observational setups, 2) model parameterizations, or 3) prior values. In what follows, we first describe how the Jacobian matrix is computed in \texttt{PICASO}. Then, we highlight some of the overall metrics that \texttt{PICASO} produces based on these Shannon information content metrics. 

\subsection{Computing the Jacobian with \texttt{PICASO}}
\label{sec:jacobian}

Computing Jacobian matrices for atmospheric models is computationally challenging because of the lack of analytical derivatives. For example, \texttt{PICASO} relies on a backend resampled opacity database that is pre-computed at discrete pressure, temperature and wavelengths, offering no continuous differentiable equation. Two main avenues exist to overcome this hurdle. The first is finite differencing. Simple finite differencing schemes include forward, backward and centered approximations. In all cases it involves running a model evaluation at either $x_j + \Delta x_j$ or $x_j - \Delta x_j$, or both in the case of the centered approach. The computational cost of finite differencing is manageable in that the maximum number of model evaluations is just 2$\times n_p$, in the centered approach or $n_p+1$ in either the forward or backward approach. Even for highly multidimensional problems this is a very small number of model evaluations, when compared to a full retrieval. Despite the computational efficiency there are still strong drawbacks to the finite differencing approach. 

For example, consider an atmospheric state described by the volume mixing ratio abundances of two gases, H$_2$O and CO$_2$. The Jacobian element for CO$_2$ computed with forward finite differencing would be: 
\begin{equation}
    K_{i,CO_2} = \frac{\mathbf{F}([CO_2 + \Delta CO_2, H_2O]) - \mathbf{F}([CO_2, H_2O])}{\Delta CO_2}.
\end{equation}
Now consider the scenario where CO$_2 <<$ H$_2$O such that there is no spectral contribution of CO$_2$ in the starting guess. We can then describe two ways in which the chosen step size for $\Delta CO_2$ fails to provide a good linear approximation of the model in a region of parameter space that is non-linear: 
\begin{enumerate}
    \item Chosen step size is too small: Because H$_2$O is the dominant molecule in the atmosphere, a step size for CO$_2$ that is too small could result in no change to the resultant spectrum because the new CO$_2$ abundance is not sufficient to outpace the H$_2$O spectral contribution. The result is a zero Jacobian vector suggesting that the atmosphere has zero model sensitivity to CO$_2$. The results of a full retrieval analysis, which usually probe a large prior range space, would not agree with this if the prior range was sufficiently large to encompass regimes where CO$_2$ contribution was visible. We emphasize that when comparing retrieval analyses and IC theory derived results it is important to choose a step size that creates a good linear approximation to the Jacobian over the expected prior range of the model.  
    \item Chosen step size is too large: conversely, choosing a step size that is too large could overestimate the Jacobian by introducing a scenario that is entirely dominated by CO$_2$. This could bias the result in the opposite direction suggesting CO$_2$ has stronger sensitivity, and a better retrieved outcome than is physically possible. 
\end{enumerate}
In both these cases we must be careful to choose a step size that provides the best possible linearization of the model over the expected prior range. Choosing a step size that is stable and that falls between these two regimes is critical to consider when using a finite differencing approach to computing a Jacobian matrix. Despite its weaknesses the only other stable avenue to computing the Jacobian is via automatic differentiation (AD). 
There are several tools, such as \texttt{JAX} \citep{jax2018github}, that have been built to assist Python users to make their code differentiable. Some substellar atmosphere codes have been built within a \texttt{JAX} framework, such as \texttt{ExoJAX} \citep{2022ApJS..258...31K} or \texttt{Exo Skryer} \citep{2026OJAp....965495L}. Although AD would provide an accurate Jacobian at a finite reference atmospheric state, it may not provide the flexibility to find a overall good linear representation of the model over a large prior range. Thus, even AD it is not a perfect solution. That being said, enabling AD for \texttt{PICASO} is an active area of development, but is not within scope for this study. Therefore, we set aside AD as a viable option to pursue for the Jacobian's computed in this work. 

Despite the drawbacks to finite differencing, it is a standard and validated approach. Perturbations are carefully chosen to be suited for the specific atmospheric state of interest, and/or are finetuned to certain benchmark cases. For example, \citet{2017AJ....153..151B} used a center finite differencing approach to compute Jacobians for a range of temperatures, atmospheric metallicities, carbon-to-oxygen ratios, and observational signal-to-noise ratios. Of these dozens of scenarios, they chose only a few cases to run a full retrieval analysis. This allowed them to pick a perturbation for their atmospheric parameters that adequately reproduced some retrieved results, ensuring a robust analysis. 

Ultimately for \texttt{PICASO}, we employ a finite differencing approach to compute the Jacobian. Users can choose between center, forward, or backward finite differencing. We offer a default perturbation of 1\%, but strongly emphasize (given the discussion here) that this value should be spot checked for robustness. Special care should be taken if users decide to use larger ($>1$\%) perturbations that have the potential to lead to nonphysical atmospheric scenarios. \texttt{PICASO} currently does not have guardrails to prevent unrealistic perturbations.   Users will also need to choose whether this perturbation should be made in logarithm or linear space. The latter consideration is often set by the convention of the \emph{a priori} covariance matrix. For example, abundances are usually given logarithm priors to cover adequate parameter space. Once users choose the scale and perturbation, we encourage spot checks with full retrieval analysis, or reproducibility of previous work (as is done in this work). If neither are possible, another strategy for users is to inspect the resultant models at the perturbed value for each model parameter. Users can check that slight variations of their chosen perturbation factor behave in a linear manner.

The last component to consider is \emph{how} the perturbation is applied to the atmospheric state within \texttt{PICASO}. In some cases, singular floating point values are trivial to perturb. For example, a cloud top pressure represents one value that can be simply increased or decreased. Vectors of floating point values are less trivial. For example, one could choose to perturb the abundance of H$_2$O at a singular value in pressure-space, versus perturbing the entire vertical extent of the atmosphere. Similar confusion applies for quantities such as the pressure-temperature profile, or surface albedo, which all are represented by a vector of floats. If the parameter that the user requests is a vector, \texttt{PICASO} will perturb the entire vector. We currently do not offer an avenue to specify a specific index within a vector. We leave this as a future code update to consider. Therefore, \texttt{PICASO} allows users to request a perturbation over any floating point number or vector that is available as input to the code. Currently this version of the code specifically computes the Jacobian for spectra. It is trivial to expand this to other functionalities of the code (e.g., climate) however it is not a focus area for this analysis.

\begin{figure*}
    \includegraphics[width=\textwidth]{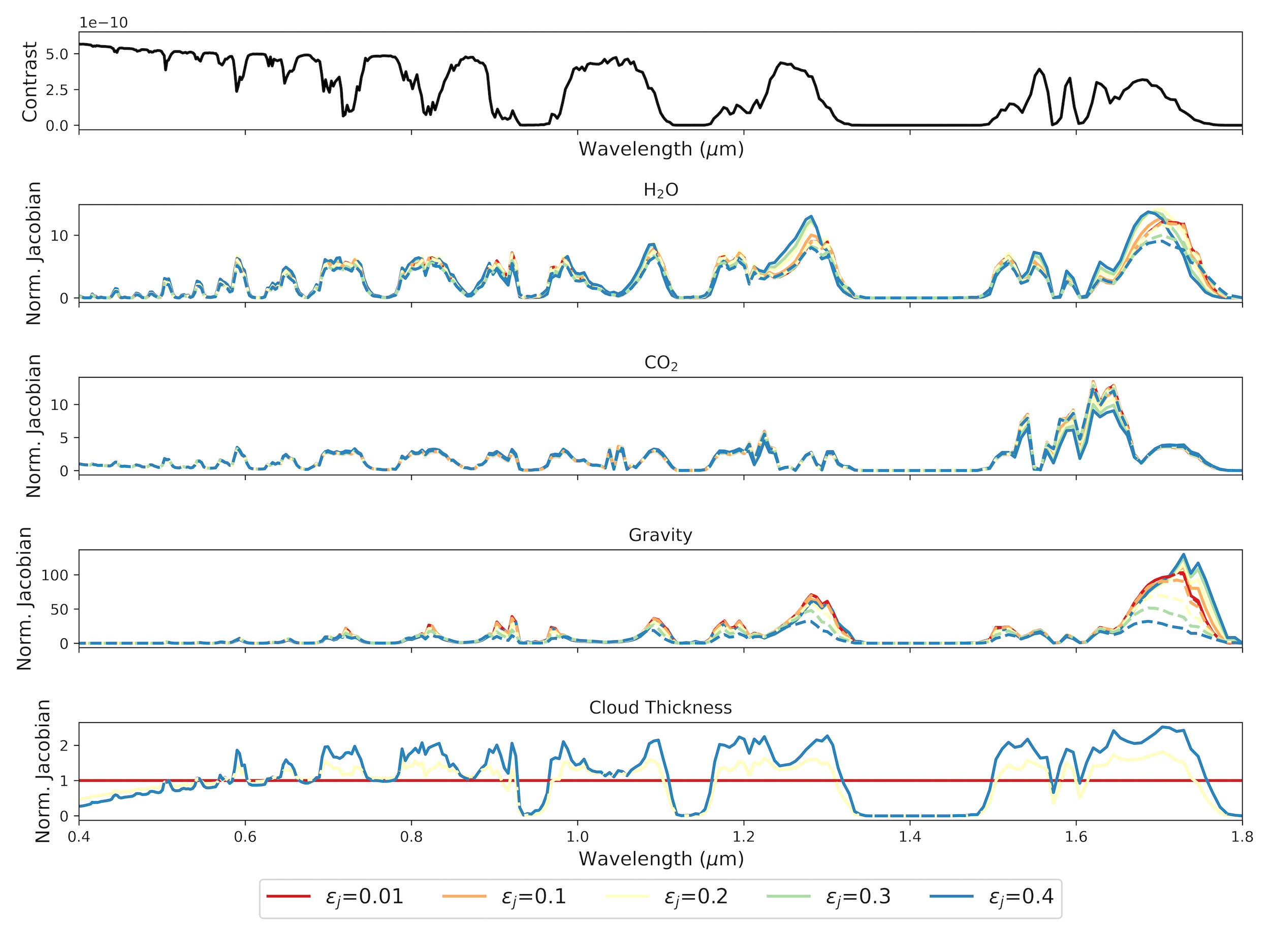}
    \caption{Example reflected light spectrum and accompanying normalized Jacobian elements  ($K_{ij} = \partial F_i / \partial x_j$), representing the derivative or the sensitivity of the spectral flux in each channel to perturbations ($\epsilon_j$) in specific model parameters. As described in \S \ref{sec:jacobian}, the case explored here is a simple reflected light example of an Earth-sized planet with an atmosphere that is equal parts H$_2$O \& CO$_2$ in an N$_2$ background. Here we show different perturbations (represented by color) and two different finite differencing methods. Dashed lines were computed via forward finite difference, and solid lines were computed via centered-finite difference. Each case has a different dependency on these input choices. For example, in the case of CO$_2$, there is little sensitivity to these choices. In the case of cloud thickness, the Jacobian is zero (value 1 red line) for the smallest perturbation. Takeaway: Jacobian calculations should always be tested against choices in derivative method and perturbation. [Code available  \href{https://github.com/natashabatalha/picaso/blob/16b428fb299bcf6663a29b828c0fdef376c6af66/docs/notebooks/K_InfoContent/IC_Stats_Tutorial.py}{\faCode}]
}
    \label{fig:jacobian_example}
\end{figure*}

Figure \ref{fig:jacobian_example} shows an example Jacobian that highlights the concepts of derivative methodology and perturbation choice, discussed here. We choose a simple example of an Earth-sized planet observed in reflected light with 10\% of both CO$_2$ and H$_2$O in an N$_2$ background with an isothermal temperature of 280~K. We include a simple gray cloud at 0.6~bar to demonstrate how the Jacobian can be computed for parameters that vary in altitude. For this example, we assume the Jacobian is composed of the abundances of H$_2$O and CO$_2$, gravity and cloud thickness only. We compute each Jacobian with \texttt{PICASO} for a range of perturbations and two different derivative methods: forward and centered. 

The Jacobian element of H$_2$O is sensitive to both these choices primarily at the band centers of the strongest H$_2$O bands. This is because these bands are saturated (fully absorbent), breaking the linearity of the model. On the other hand CO$_2$ has no regions of fully saturated bands and therefore the choice of derivative or perturbation size does not largely affect the Jacobian. Gravity and cloud thickness, which both affect the continuum, are the most sensitive to these choices. For example, for the cloud thickness the smallest perturbation explored ($\Delta x_j=x_j*\epsilon_j$ where $\epsilon_j$=0.01,red) results in a zero-valued Jacobian (in the plot it is normalized to 1) because the perturbation is not large enough to move the cloud deck into the neighboring pressure-level of the grid used. In this case, the discontinuity in the parameterization of this specific model setup makes linearization difficult and blindly choosing a perturbation would drastically alter results. Ultimately, we emphasize the need for spot checking perturbations and finite differencing methods to ensure the final results are not sensitive to these choices. \texttt{PICASO} provides the necessary flexibility to do so rapidly. 

The full tutorial on computing Jacobians with \texttt{PICASO} is available here ( \href{https://github.com/natashabatalha/picaso/blob/16b428fb299bcf6663a29b828c0fdef376c6af66/docs/notebooks/K_InfoContent/IC_Stats_Tutorial.py}{\faCode}). The main function that computes the Jacobian in \texttt{PICASO} is available here ( \href{https://github.com/natashabatalha/picaso/blob/16b428fb299bcf6663a29b828c0fdef376c6af66/picaso/information_content.py\#L85}{\faCode}).

\begin{figure*}
    \centering
    \begin{subfigure}[b]{0.48\textwidth}   
        \centering
        \includegraphics[width=\linewidth]{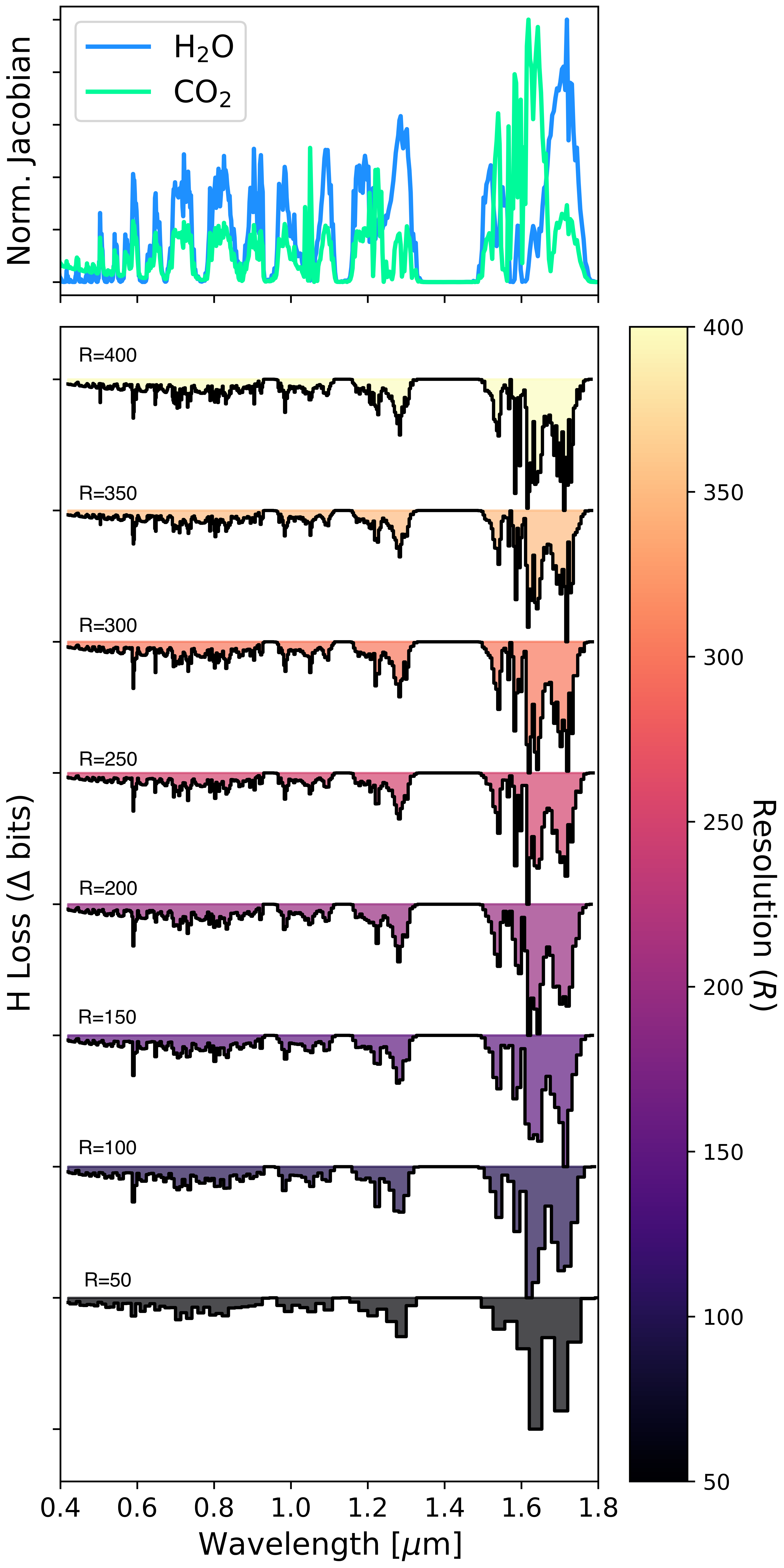}
        \caption{Information loss by wavelength considering all model parameters}
        \label{fig:loss_H}
    \end{subfigure}
    \begin{subfigure}[b]{0.48\textwidth}   
        \centering
        \includegraphics[width=\linewidth]{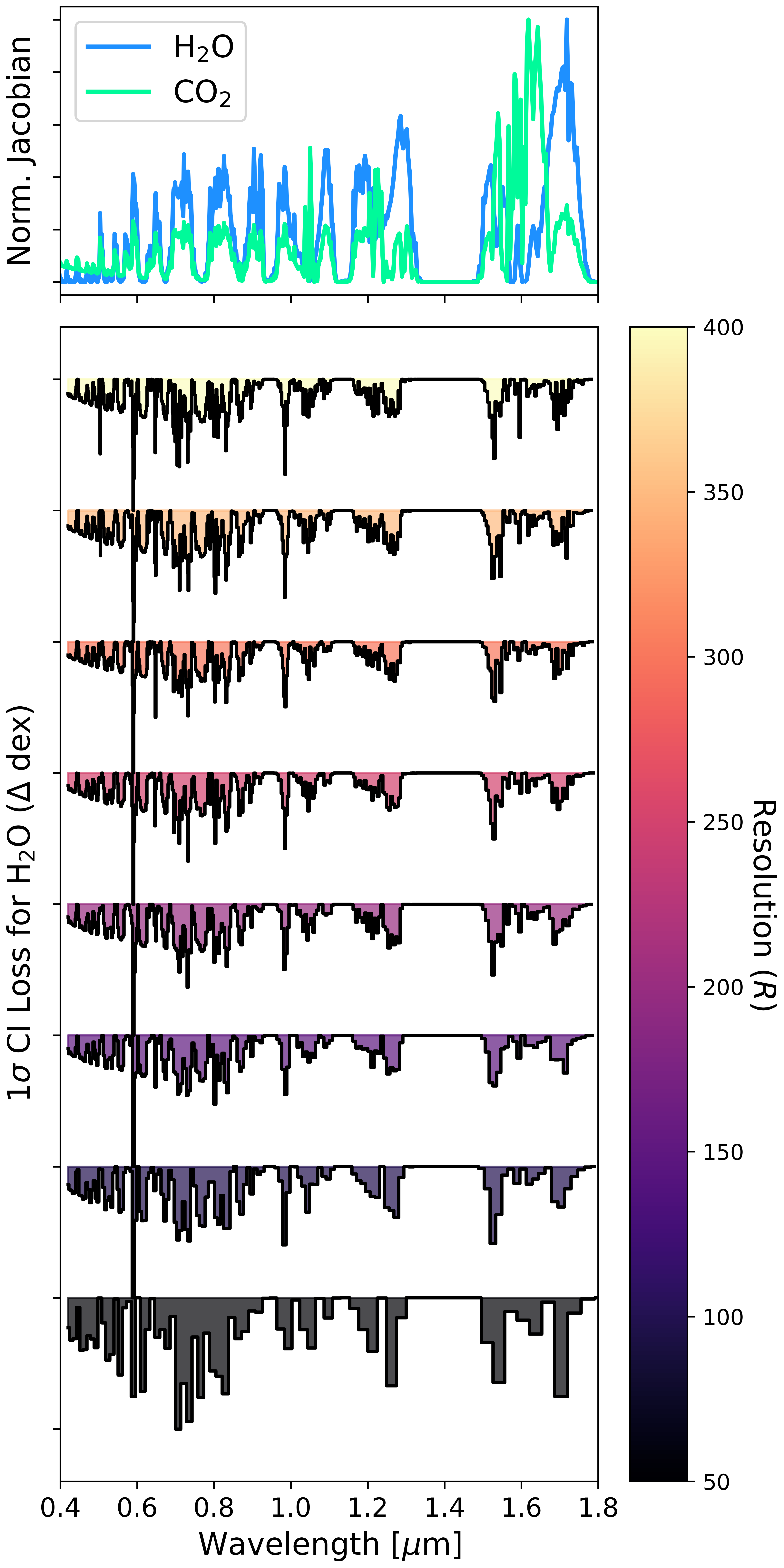}
        \caption{Information loss by wavelength considering only H$_2$O precision}
        \label{fig:loss_h2o}
    \end{subfigure}
    \caption{Top panel for a) and b): A reference Jacobian for H$_2$O and CO$_2$ identical to that shown in \ref{fig:jacobian_example} computed via center-finite differencing with a perturbation $\epsilon_j=0.1$. The Jacobian vectors are shown at R=400. Bottom panels: The loss of information induced by the removal of a certain size resolution element (shown via the color bar on the left). In (a) ``information'' is the information content, $H$ of the system. In (b) the ``information'' is the 1$\sigma$ H$_2$O constraint interval. By cross-referencing the information loss to the Jacobian vectors we start to understand where critical loss regions are in resolution space. For example, the looking at (a) CO$_2$ parameter sensitivity centered between H$_2$O at 1.2 and 1.3$\mu$m is consolidated to one data point at R=100 and disappears at R=50. Comparing (a) and (b) we quickly see how H$_2$O information is weighted toward the optical whereas information of the system is weighted toward the infrared. Takeaway: IC metrics can help focus retrieval efforts by highlighting important regions of wavelength, or for example, information loss cliffs in resolution space. [Code available  \href{https://github.com/natashabatalha/picaso/blob/16b428fb299bcf6663a29b828c0fdef376c6af66/docs/notebooks/K_InfoContent/IC_Stats_Tutorial.py}{\faCode}] }
    \label{fig:ic_loss_ex}
\end{figure*}

\subsection{Comparative Shannon IC Metrics}
\label{sec:icmetrics}
In \S\ref{sec:methods} we gave an overview of the overall Shannon information content metrics. Here we show how to utilize those metrics in a comparative framework by showcasing an example of information loss across wavelengths. We build upon the Jacobian example shown in Figure \ref{fig:jacobian_example}. We assume large uninformative priors for the four parameters in the atmospheric vector. For measurement error, we assume an albedo signal-to-noise ratio, SNR=10, defined at $\lambda=0.55\mu$m. Then, for a range of resolutions, R=50--400, we compute information loss as a function of wavelength via \texttt{PICASO} function, \texttt{loss\_by\_wave} \href{https://github.com/natashabatalha/picaso/blob/16b428fb299bcf6663a29b828c0fdef376c6af66/picaso/information_content.py\#L440}{\color{gray}\faCode}. This function computes the total information content, $H$, and the posterior covariance parameters (Eqns. \ref{eqn:H}, and \ref{eqn:s_hat}). Then it repeats this process while sequentially removing one wavelength element from the Jacobian, and associated measurement error matrix. The result, (e.g., $\Delta H=H-H_{-i}$) showcases one of the many utilities of information content theory. 

First, let's evaluate  Figure \ref{fig:loss_H}, which considers the loss of information content, $H$, when a single resolving element is removed. We highlight two of several conclusions that can be drawn from this exercise. By assessing the information loss across the shown parameter space (0.4--1.8$\mu$m) it is clear that the information rich area  exists from 1.6--1.8$\mu$m. The region centered at 1.4$\mu$m is contributing no new information to the system. Before full retrievals are conducted these IC-based metrics can help refine what regions of wavelength parameter space are most important to focus on. Additionally, by comparing the information loss as a function of resolution we begin to understand where ``information cliffs'' might exist. For example, from 1.5--1.8$\mu$m the Jacobian tells us there is parameter sensitivity to both H$_2$O and CO$_2$. Toward high resolution (yellow, R=400) individual lines of  H$_2$O are CO$_2$ differentiable. Toward lower resolution (black, R=50) there is only one singular resolution element representing H$_2$O and CO$_2$, respectively.

Next, Figure \ref{fig:loss_h2o}, considers the loss of information through evaluating how the 1$\sigma$ constraint interval of H$_2$O predicted from the covariance matrix decreases with the removal of a single resolving element. Here it is interesting to note that the information rich area is more heavily weighted toward the optical, rather than the infrared as was shown in the $H$ loss figure \ref{fig:loss_H}. These exercises enable the prioritization of wavelength regions, especially in how it pertains to the sequencing of an observation.  Ultimately, we advocate for these types of metrics to be created prior to full retrieval exploration to understand parameter sensitivity. 


\subsection{Singular Value Decomposition}
\label{sec:SVD} 
Visual inspection of the components of the Jacobian can be informative in identifying degeneracies between parameters. Regions of wavelength space where the Jacobian vectors for two model parameters seem to be correlated, could represent regions of degeneracy. However, simply inspecting the correlations between Jacobian columns does not enable the measurement error to be considered. The off-diagonal elements of the posterior covariance matrix ($\mathbf{\hat{S}}$), which does consider both the measurement error and the prior, is one way to quantitatively inspect parameter degeneracies. Another strategy that examines the specific geometry of a matrix is singular value decomposition (SVD). SVD breaks a matrix down into fundamental geometric components. Conducting SVD on the Fisher Information Matrix (FIM=$\mathbf{K_a}^\mathrm{T} \mathbf{S_e}^{-1} \mathbf{K_a}$), is the last metric which our suite of tools focuses on. 

If we conduct SVD on the FIM we are effectively conducting an eigenvalue decomposition to understand the geometry of the matrix. We use \texttt{numpy.linalg.svd} (\href{https://github.com/natashabatalha/picaso/blob/62671146ab93d1a81661b81f11535a532c64ff2b/picaso/information_content.py\#L365}{\faCode}) which breaks down the FIM into three components: 
\begin{equation}
    F = U \cdot S \cdot V^H
\end{equation}
The two outputs which provide the information needed to assess degeneracies between parameters are $V^H$, the parameter combinations, and $S$, the information strength. $V^H$ represents the principal axes of information. Each row is a vector element corresponding to a model parameter in the Jacobian and contains a quantity that defines the linear combination of the parameters. When one or more values of $V^H$ is high it suggests that the corresponding model parameters are coupled. $S$ contains the singular values themselves and represents the curvature of the likelihood surface. High values of $S$ represent steep directionality and the presence of information. Low values of $S$ represent flat directionality and the presence of little information. The strategy in degeneracy assessment is to put these two quantities together by finding the smallest value of $S$, the corresponding row in $V^H$, and the elements in that row which contain the largest coefficients. Visually, it is easier to conceptualize this quantity. We demonstrate the use of SVD in assessing parameter degeneracies in \S \ref{sec:feng}.

\section{Spectral Sensitivity Analysis of Earth's Reflected Light Spectrum}
\label{sec:feng}
Even prior to the conceptualization of the Habitable Worlds Observatory, there was rich literature surrounding what observational requirements were needed to observe an Earth-like exoplanet via reflected light spectroscopy. Thus far, nearly all have utilized retrieval methodologies. For example, \citet{2018AJ....155..200F} was one of the first studies which evaluated what information could be retrieved from 0.4--1$\mu$m spectra at spectral resolutions of R=70 and R=140, and at signal-to-noise ratios (SNR) ranging from 5--20. This initial study focused on a validated Earth model comprised of 11 total atmospheric parameters: 1) 3 total describing molecular abundances (ozone, oxygen, and water),  2) 2 total describing planet properties (planet radius and surface gravity), 3) 2 total describing surface properties (surface pressure and surface albedo), and 4) 4 total describing cloud properties (cloud top pressure, cloud optical depth, cloud thickness, and cloud coverage fraction). Many key degeneracies were noted in this work such as those between surface pressure and surface albedo, surface pressure and gravity, planet radius and surface albedo. As pointed out, these degeneracies are directly tied to physical spectral properties such as the influence of both gravity and surface pressure on column mass. \citet{2018AJ....155..200F} also computed the complex posterior probability distributions of each parameter. For example, they showed how even for SNR=20 and R=140 data, only limits could be placed on most cloud parameters.

Since the work of \citet{2018AJ....155..200F} there have been numerous other retrieval-based works more specifically related to HWO with the goal of understanding the observational parameters required to constrain atmospheric parameters on an Earth-like planet. These works have come from independent frameworks such as the BARBIE framework \citep{2023AJ....166..129L, 2024AJ....167...27L, 2025AJ....169...50L}, the \texttt{rfast} modeling suite \citep{2024ApJ...969L..22S, 2025ApJ...995..173S, 2025arXiv250714771K}, and the \texttt{EXOREL} modeling suite \citep{2020AJ....159..175D, 2022AJ....163..299D, 2023AJ....166..157D, 2025AJ....169...97D}. Each has tackled complex problems relating to the formulation of mission requirements for HWO. Of these studies, the simplest modeling setup to reproduce is that of the original work done by \citet{2018AJ....155..200F}. 

Specifically, the legacy model used in \citet{2018AJ....155..200F},  which was originally developed by \citep{1989Icar...80...23M} and updated by \citep{2010ApJ...724..189C}, is the identical legacy code which formed the foundation of the original \texttt{PICASO} reflected light model \citep{2019ApJ...878...70B}. Additionally, \citet{2018AJ....155..200F} provides a comprehensive table of both upper and lower 1-$\sigma$ constraint intervals for which we can directly compute our IC theory statistics. We note, however, that our goal here is not to provide a deep quantitative model comparison/reproduction of that provided in \citet{2018AJ....155..200F}. Our goal is only to emphasize how Linear-Gaussian approximated 1$\sigma$ constraint intervals compares to those derived from retrievals. Therefore, below we describe the model setup we use to roughly reproduce the model of \citet{2018AJ....155..200F} and show the resulting Jacobian. Next we provide a comparison of the retrieved statistics. Lastly, we showcase the utility of the other statistics that \texttt{PICASO} provides and describe how they relate to conclusions that are also drawn from more time intensive retrieval studies. 

\begin{figure*}
    \includegraphics[width=\textwidth]{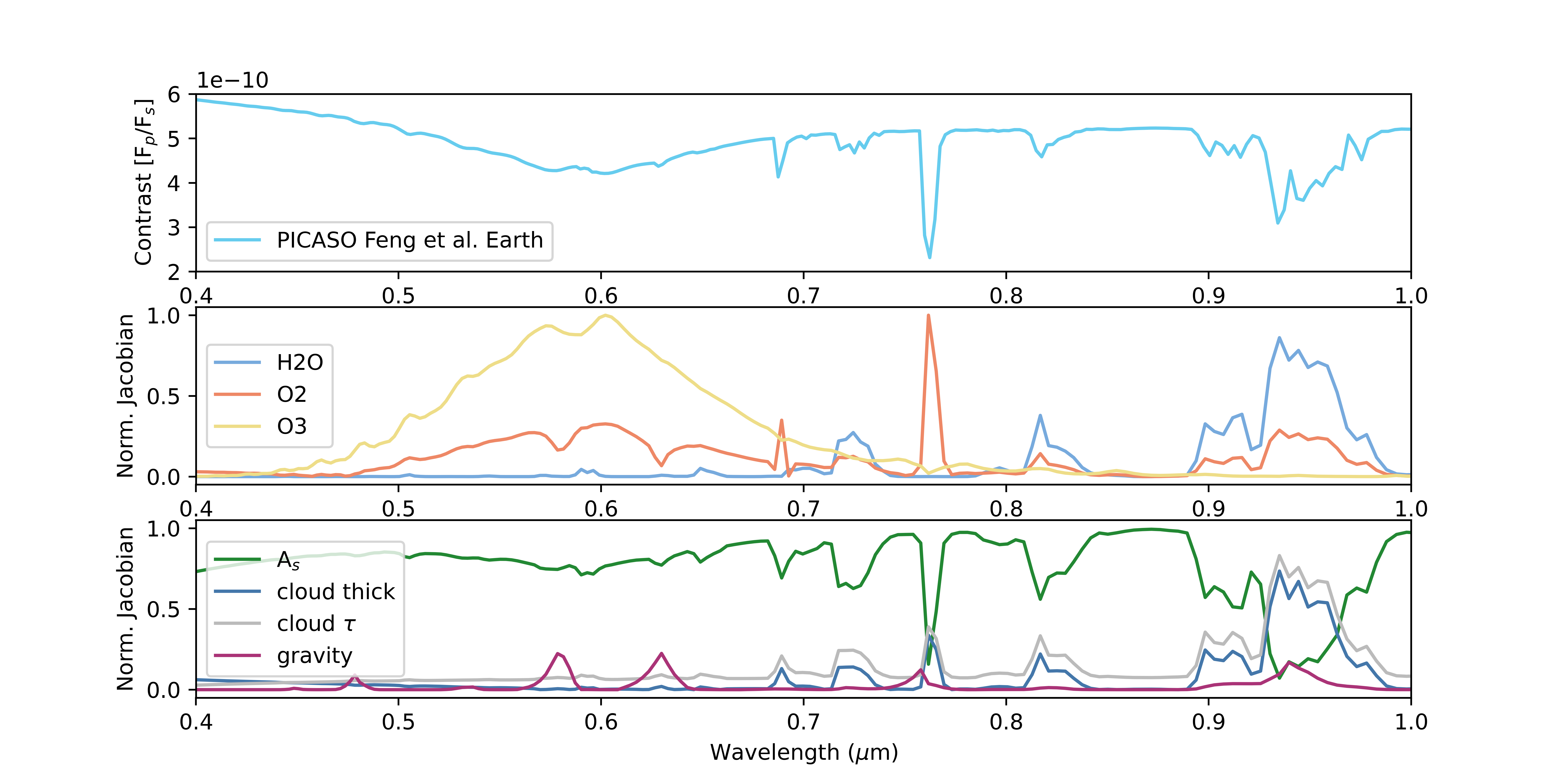}
    \caption{Top panel: \texttt{PICASO} reproduction of the Earth-like atmosphere model from \citet{2018AJ....155..200F}, demonstrating consistent spectral morphology and contrast levels. Middle and bottom panels: Normalized Jacobian elements ($K_{ij} = \partial F_i / \partial x_j$), representing the derivative or the sensitivity of the spectral flux in each channel to perturbations in specific model parameters. Middle panel: Jacobians for molecular abundances; high-sensitivity regions (peaks) directly map to the discrete molecular absorption features seen in the top panel. Bottom panel: Jacobians for continuum-driving parameters: surface albedo ($A_s$), cloud thickness, cloud optical depth ($\tau$), and gravity. The results highlight the anti-correlated nature of surface albedo and cloud properties, which can create parameter degeneracies during a retrieval. For gravity, sensitivity peaks correspond to regions of collision-induced opacity (e.g., $O_2$-$O_2$ features blue- and red-ward of 0.6 $\mu$m). Takeaway: These Jacobian elements quantify the sensitivity of the forward model to each parameter, providing the fundamental mathematical basis for calculating information content theory metrics.
}
    \label{fig:jacobian}
\end{figure*}

\subsection{The Initial Model and Computed Jacobian}
\label{sec:bench_jac}

We replicate the model presented in \citet{2018AJ....155..200F} by using their input parameters listed in their Table 4 and 5. We opt for as many identical model choices as possible. We use the same 1) assumption of isothermal pressure-temperature profile at T=250~K, 2) two-term Henyey-Greenstein phase function weighted by the Rayleigh contribution to treat the direct scattering beam from clouds, 3) implementation of patchy clouds (i.e., $f_c \times$ cloudy +  $(1-f_c) \times$ cloud free, where $f_c$ is the cloudiness fraction), 4) removal of any Raman scattering or polarization, 4) cloud asymmetry factor ($\bar{g}=0.85$) for a single grey H$_2$O cloud deck. Many of these additional modeling choices represent parameters that are not explicitly included the Jacobian calculation but are needed to create a reflected light spectrum of an exoplanet. The differences in modeling choices between \texttt{PICASO} and the \citet{2018AJ....155..200F} include: 1) updated opacities computed from \citet{wogan_2025_17381172} accounting for latest HITRAN data release \citep{GORDON2026109807}, 2) single scattering albedo ($\bar{\omega}=0.97$ vs. 1 in \citet{2018AJ....155..200F}), 3) cloud optical depth and pressure regridding. For \#1, our spectral comparison shows very similar feature morphology to \citet{2018AJ....155..200F} suggesting that this is not expected to drive major differences between our analyses. For \#2, \citet{2018AJ....155..200F} assumed a purely scattering value of $\bar{\omega}=1$, which would result in \texttt{NaN} in \texttt{PICASO}'s implementation of the \citet{1989JGR....9416287T} radiative transfer scheme. The single scattering sets the overall brightness of the spectrum. Opting for a value near 1, $\bar{\omega}=0.999$, resulted in a contrast (flux of the planet / flux of the star) that was too high ($\sim$9.4e-10 vs. $\sim$5.5e-10 at $0.4\mu$m). The 3D ocean model shown in \citet{2018AJ....155..200F} had a contrast of $\sim$7e-10 at $0.4\mu$m. Given the simplicity of the grey cloud model in these analyses, we opt for a value that generally reproduces the continuum baseline of their reported spectrum. It is also possible that \#3 contributed to the need for slight differences in single scattering albedo. \texttt{PICASO}'s implementation of a grey cloud simply sets an extinction per layer. We choose an optical depth=5 for each of the layers within the grey deck. \citet{2018AJ....155..200F} set the optical depth for each layer to be $\tau/N_c$ where $N_c$ is a variable number of cloud layers proposed by their retrieval algorithm. Despite these minor differences, the \texttt{PICASO} spectrum (shown in Figure \ref{fig:jacobian} top panel) reproduces all major spectral feature morphologies shown in Figure 4 of \citet{2018AJ....155..200F}.

With the initial atmospheric state set, we can move to the calculation of the Jacobian. We again include all the same parameters in our calculation of the Jacobian. The one minor difference is that rather than include both gravity and radius, we only include gravity as a Jacobian parameter. This is because \texttt{PICASO}'s input specifications will override a gravity input if the combination of mass and radius are given as input. Therefore, perturbing gravity alone, while also providing mass and radius, would result in an artificially zero Jacobian vector. As shown in the following sections, this does not affect our ability to reproduce the results. 

We use a center-finite differencing approach to compute the Jacobian. To determine the perturbation, we first use the \texttt{PICASO} suggested default of $\epsilon_j=$0.01. All of the Jacobian parameters besides gravity, cloud thickness, and cloud optical depth showcased stability at this level and were not strongly affected by slight variations in this value. Cloud thickness and cloud optical depth required higher perturbation values in order to yield a better overall linearized approximation of the model over a wide parameter space for these parameters. Similar to the example shown in Figure \ref{fig:jacobian_example} we had to choose a cloud thickness perturbation value that was large enough to kick the cloud into a neighboring pressure-grid value (0.05). Cloud optical depth, similarly, had to be perturbed by a value that was large enough to exit our initial assumption of an optically thick cloud (initial state had $\tau=5$). A perturbation by $\epsilon_j=$0.2  was enough to create a non-zero Jacobian. Similarly the gravity required a larger perturbation (0.2) to affect the column density of the spectrum when specifying a fixed surface pressure. Ultimately when comparing IC statistics with a retrieval, these choices are driven by the need to create a linear approximation of the Jacobian over the prior parameter space that the retrieval would explore. 

The resultant Jacobian of key parameters are shown in the bottom two panels of Figure \ref{fig:jacobian}. We note that all of the Jacobian vectors are normalized to their absolute magnitudes for visual clarity. The middle panel shows the model sensitivity to specific abundances of gases. The bottom panel shows a representative set of parameters that affect the reflected light spectrum's continuum. Comparing the Jacobian elements of the gas abundances to the modeled spectrum, we can easily identify each molecule's contribution to the overall spectrum. The continuum parameters are less intuitive, when compared to the overall spectrum. However, key insights can be gained as well. First, it is interesting to note the anti-correlation between the surface albedo and the cloud properties. As expected, when the model sensitivity to the cloud increases, the model sensitivity to the surface decreases. We also see that the Jacobian elements of the cloud thickness and optical depth have nearly identical morphologies because they both contribute to the overall brightness of the spectrum. They only differ slightly near 0.4$\mu$m where the cloud thickness contains relatively more information because of its interference with the Rayleigh scattering slope. Lastly, the Jacobian for gravity is particularly interesting. It exhibits peaks that do not map to large molecular bands shown in the abundance Jacobian Figure. This is because the Jacobian of gravity is sensitive to collision induced opacity, which can be seen as smaller features in the full spectrum. Ultimately, the behavior of all the Jacobian vectors highlight fundamental parameter sensitivities expected from basic atmospheric physics. 

\begin{figure*}
    \includegraphics[width=0.95\textwidth]{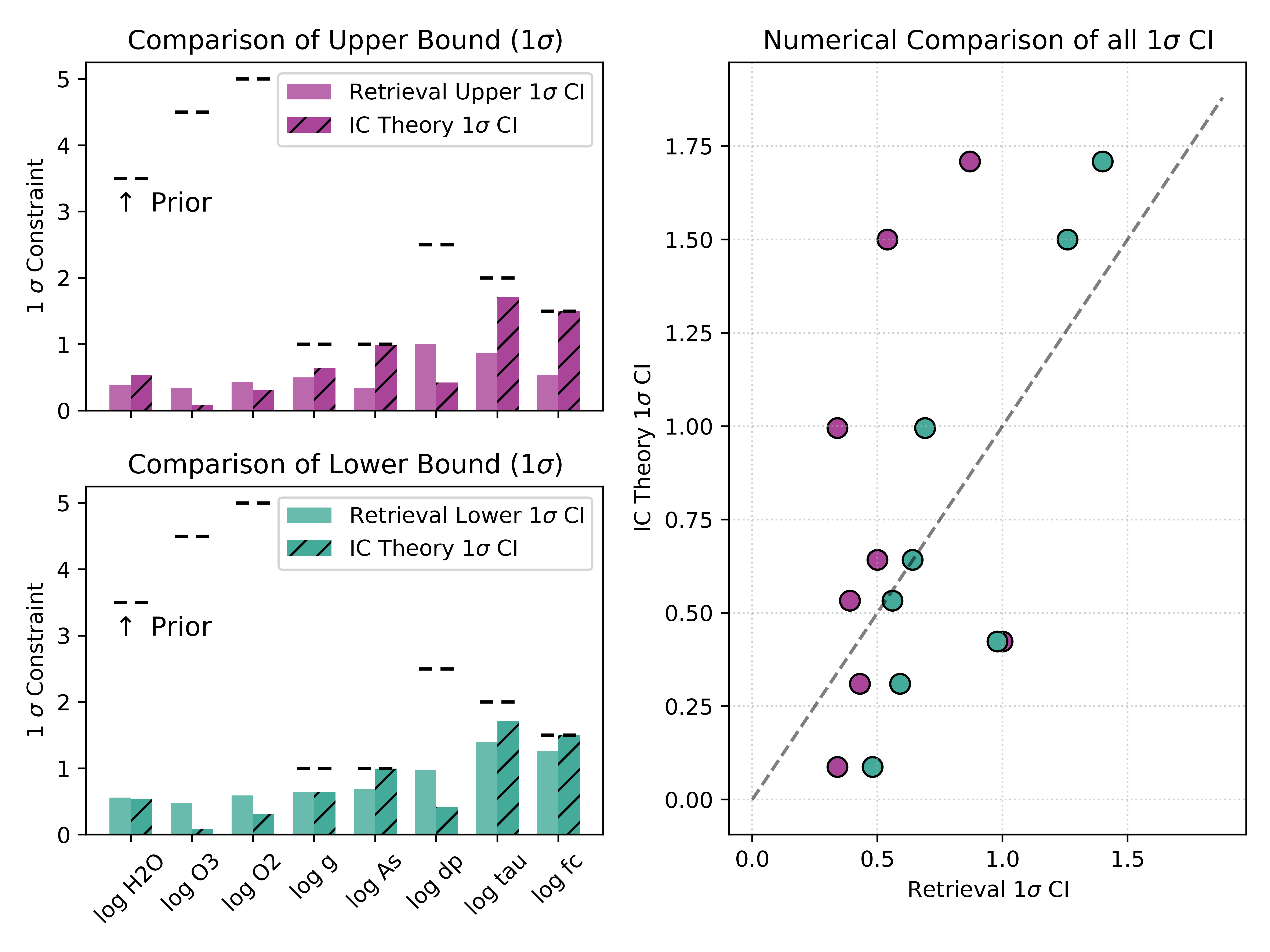}
    \caption{Left panels: Comparison of upper and lower bounded 1$\sigma$ constraint intervals (CIs) for the $R=140$/SNR=20 case in \citet{2018AJ....155..200F} (solid bars) against those predicted by IC theory (hashed bars). IC theory assumes fully Gaussian posterior probability distributions, where the 1$\sigma$ uncertainties are determined by the diagonal elements of the posterior covariance matrix ($\hat{S}$). Dashed lines indicate the assumed prior for each parameter; CIs that approach these values indicate model parameters that are unconstrained by the data. Right panel: A quantitative comparison of the retrieval uncertainties derived from both methods. Takeaways: 1) IC theory demonstrates robust performance, reproducing all lower bound CIs within 0.5 dex. 2) The most significant disagreements ($\sim$1 dex) arise for parameters whose retrieved posterior probability distributions are not Gaussian.}
\label{fig:CI}
\end{figure*}

\subsection{Direct Comparison of Retrieval and IC Theory-derived 1$\sigma$ Constraint Intervals}

\begin{figure*}
    \includegraphics[width=0.75\textwidth]{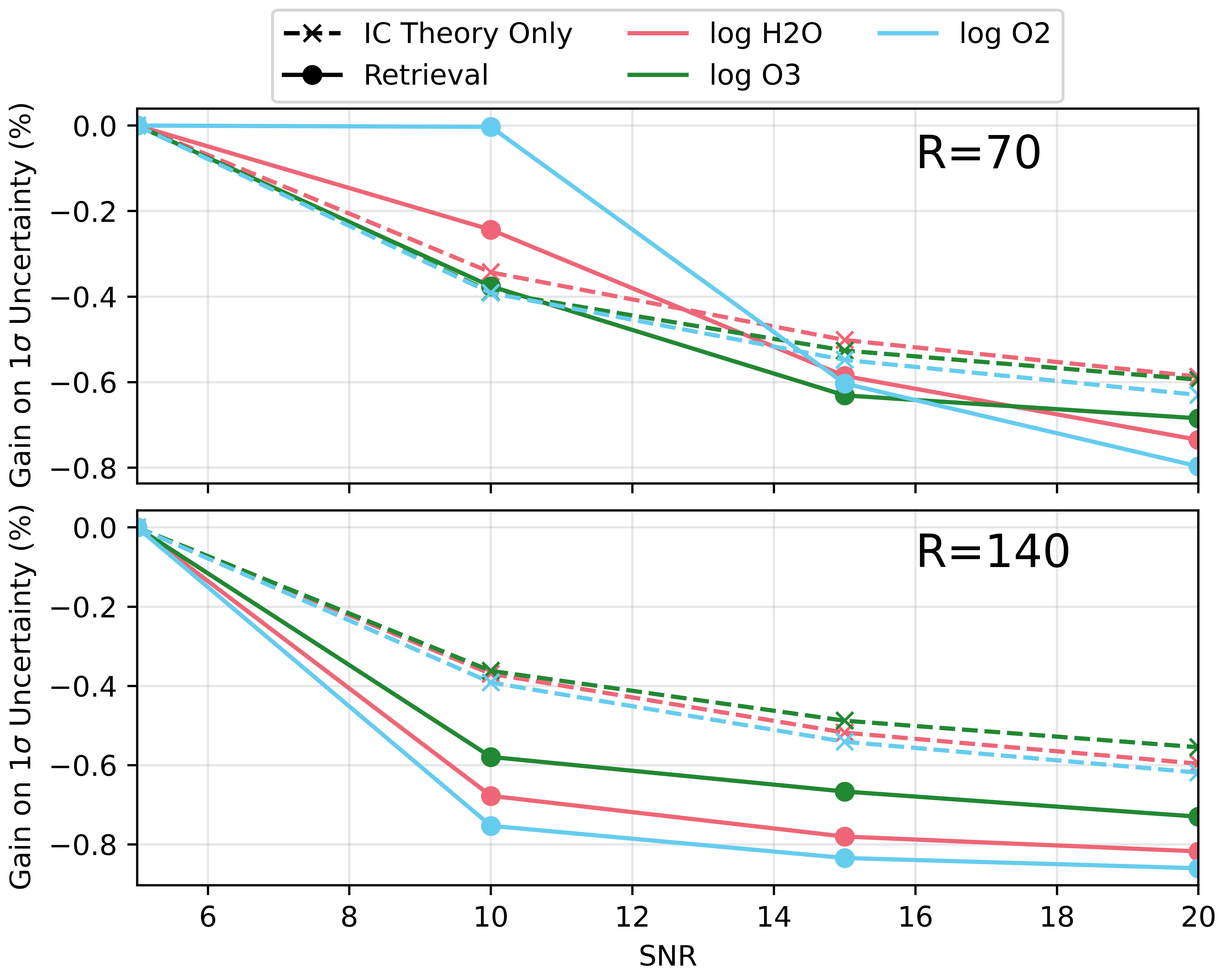}
    \caption{Building on the comparison of absolute constraint magnitudes in Figure \ref{fig:CI}, this figure isolates the 1$\sigma$ constraint intervals (CIs) for molecular abundance parameters to demonstrate the capacity of IC theory to quantify information loss as a function of degrading signal-to-noise ratio (SNR). The percent gain is normalized to the 1$\sigma$ uncertainty at SNR=5 in order to emphasize the comparison of the loss with SNR (absolute error deviations are shown in Figure \ref{fig:CI}). Takeaways: IC theory disagrees with retrievals the most in the low-SNR regime, where non-Gaussian posterior distributions  dominate. Ultimately there is robust agreement between the per-parameter information loss reported by full retrievals and those predicted by IC theory. 
}
\label{fig:CI_loss}
\end{figure*}

Given the computed Jacobian discussed in \S \ref{sec:bench_jac} we can now proceed with computing other spectral sensitivity analysis metrics. The two additional components needed are the error covariance vector, and the prior matrix. \citet{2018AJ....155..200F} computed an SNR scaling (their Figure 6). For the retrieval cases most pertinent to HWO, they explored two spectral resolution cases (R=70, R=140) at four different SNRs (=5,10,15,20) which they define at $0.55\mu$m. We adopt this scaling to compute error covariance vectors for the 8 total cases. We adopt the prior values from their Table 1. Note that within the IC theory framework we do not specify a range, only the 1$\sigma$ Gaussian width of the prior region.

Figure \ref{fig:CI} shows the R=140 numerical comparison of the 1$\sigma$ constraint intervals (CIs) reported in \citet{2018AJ....155..200F} and the Shannon information content derived posterior covariances for a representative set of atmospheric parameters. Because IC theory assumes Gaussianity, we compare both the upper and the lower bounds independently. As expected, in most cases, when a Linear-Gaussian approximation is applied the metrics yields  more favorable outcomes when comparing the numerically derived constraints. This is why we have emphasized the use of IC theory alongside full retrieval analyses. That being said, IC theory demonstrate robust performance, reproducing all lower bound CIs within 0.5 dex. The most significant areas of disagreement ($\sim$1 dex) arise for parameters with high levels of parameter degeneracy (for example, cloud thickness and optical depth), where the actual posterior probability distribution is expected to deviate from a Gaussian. The degeneracy between cloud thickness and optical depth was demonstrated qualitatively by assessing their Jacobian elements in Figure \ref{fig:jacobian}. There are two instances (for surface reflectivity and cloud fraction) where IC theory suggests a fully prior-dominated regime whereas the retrieval demonstrates a slight deviation from the prior. In these cases, an assessment of the upper vs. lower bounds of the parameter shows that in the retrieval only a limit is being placed on the value. Therefore, those parameters are also themselves prior-dominated in the retrieval. Though quantitative agreement does not map identically, qualitatively, the IC theory derived 1$\sigma$ constraints can be used to guide regions of parameter space where full retrieval studies should be conducted. 

In addition to posterior covariance matrix elements, we can also leverage the \citet{2018AJ....155..200F} results to demonstrate how IC theory metrics compare to full retrieval outcomes in terms of how they quantify information loss (or gain) with varying R and SNRs. The results of this analysis are shown in Figure \ref{fig:CI_loss} for both R=70 and R=140. IC theory exhibits its highest degree of disagreement with full retrievals in the low-SNR regime (e.g., the transition from SNR=10 to SNR=5). In these instances, poor data quality leaves many model parameters unconstrained by the observation, leading to non-Gaussian posterior distributions that violate IC theory’s fundamental Gaussian assumptions. Overall, however, there is robust agreement between the per-parameter information loss reported by full retrievals and those predicted by IC theory. For example, at R=140 both techniques suggest little improvement on all constraint intervals when the data quality is improved from SNR=10 to SNR=20. This validates the use of Gaussian-Linear metrics as a computationally efficient diagnostic tool for rapidly identifying optimal observing modes and conducting instrument trade-off studies. 

\subsection{Degeneracy Assessment: Application of Singular Value Decomposition}

\begin{figure*}
    \includegraphics[width=0.75\textwidth]{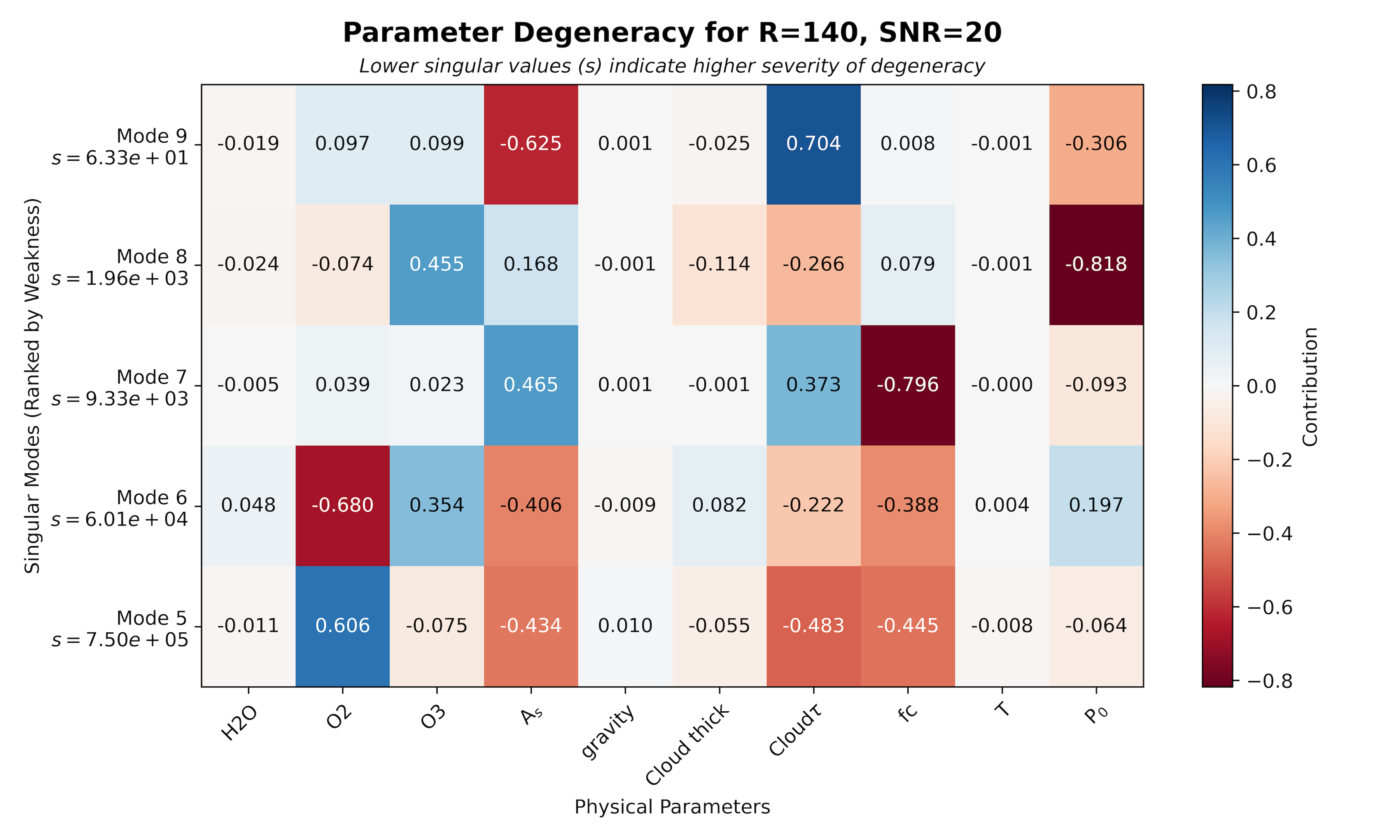}
    \caption{Singular value decomposition (SVD) analysis of the Fisher information matrix for the Jacobian shown in Figure \ref{fig:jacobian} and the R=140, SNR=20 observation  setup from \citet{2018AJ....155..200F}. Degeneracy assessment using SVD involves visualizing $V^H$, representing parameter combinations, and $S$, the information strength. The heatmap displays the composition of the singular modes (rows), ranked by their corresponding singular values ($S$) from smallest to largest. Each cell represents the contribution of a physical parameter (columns), as defined by $V^H$. Lower values of $S$ indicate higher degree of degeneracy. In those regions of low $S$, parameters with high components of $V^H$, represent the parameters that are degenerate. For example, here information content identifies surface albedo, cloud optical depth, and surface pressure to be the most degenerate parameters in this atmospheric model.  Takeaway: Degeneracies between atmospheric parameters can be easily identified with SVD analysis of the Fisher information matrix. 
 ).
}
\label{fig:SVD}
\end{figure*}

Gaussian-Linear approximations, which we refer to as IC theory,  broadly reproduce $1\sigma$ constraint intervals derived from full retrieval analyses. However, disagreement begins to appear in regions where retrieval-derived posterior distributions are likely non-Gaussian. \citet{2018AJ....155..200F} identified many scenarios in which physical parameters were degenerate or even fully unconstrained. Therefore, this presents a useful test case to determine whether or not the directionality derived from the single value decomposition (SVD) of the linear-Gaussian Fisher matrix can qualitative identify the same degeneracies found in the full retrieval analysis of \citet{2018AJ....155..200F}.  As a reminder, in SVD we are assessing the local curvature of a matrix and are interested in visualizing $S$ and $V^H$, which combined give us a rank ordering of parameter degeneracies for a given system. Figure \ref{fig:SVD} shows both quantities. The rows of the heat map correspond to the rank order of $S$ (with the top row corresponding the lowest values of $S$). Each column contains the corresponding value of $V^H$, which indicates specific weights (0--1) of each of the physical parameters. Degenerate parameters are those with large-magnitude weights in the singular vectors associated with the smallest singular values (i.e., the top rows). 
From the first row in Figure \ref{fig:SVD}, surface albedo, cloud optical depth, and surface pressure have the most severe degeneracy. The second row (corresponding to a larger singular value), have large $V^H$ components for surface albedo, cloud optical depth, cloud fraction, and surface pressure, further validating their parameter confusion. These same degeneracies are apparent in the posterior probability distributions derived from the atmospheric retrievals in \citet{2018AJ....155..200F}, suggesting a qualitative agreement between the SVD-derived degeneracy directions and the retrieval results. Ultimately, this provides evidence that the local linear-Gaussian Fisher approximation can be used to  diagnose dominant degeneracy directions before computational intensive retrievals are run.

\section{Discussion \& Conclusions}
\label{sec:d_and_c}
We have presented a software update to the open-source radiative transfer tool \texttt{PICASO}, which integrates a comprehensive suite of spectral sensitivity analysis tools. At the core of the toolkit is a flexible approach to locally linearize the forward model and compute a Jacobian matrix. A Jacobian which quantifies the sensitivity of a spectral model to any set of valid \texttt{PICASO} atmospheric parameters. Next, we provide users with a suite of statistical metrics to assess spectral sensitivity and degeneracies. This includes  information content, degrees of freedom, posterior covariance estimates, and singular value decomposition for degeneracy assessment.
We demonstrate the toolkit's utility through a benchmark study of an Earth-like planet in reflected light, following the legacy work of \citet{2018AJ....155..200F}. Our primary conclusions are as follows:

\begin{enumerate}
\item Computational Efficiency: Unlike full retrieval methods that require hundreds of thousands of model evaluations, linear-Gaussian approximations of the forward model require a maximum of only $2 \times n_p$ evaluations to compute the Jacobian. Once the Jacobian is established, trade-off studies involving varying spectral resolutions, signal-to-noise ratios, and wavelength coverages can be performed instantaneously. We demonstrate key metrics and diagnostics that can be created rapidly.
\item 	Robustness of Metrics: Fisher-matrix-based estimates of 1$\sigma$ constraint intervals can reproduce those derived from full Bayesian retrievals. For posterior distributions that are expected to be Gaussian, linear-Gaussian estimates agree with retrieval results within 0.5 dex. At low SNR, prior dominated regimes where non-Gaussian behavior dominates IC begins to deviate. Therefore, for high-priority science cases, linear-Gaussian approximated values should always be validated against full retrievals. 
\item Enabling Assessment of Information Loss: We demonstrate the strength of linear-Gaussian approximations in their ability to perform comparative assessments. We show examples where we can  quickly identify critical wavelength regions and resolution thresholds where information gain is maximized or where ``information cliffs'' occur -- points where degrading data quality leads to a total loss of parameter sensitivity. We find these conclusions to be robust and hope they can be used to focus full retrieval efforts.  
\item Diagnostic Power for Degeneracies: The Jacobian, averaging kernel matrices, and singular value decomposition provide immediate physical insight into parameter degeneracies, such as the anti-correlation between surface albedo and cloud properties. These diagnostics can pre-identify non-Gaussian behaviors that would typically require time-intensive retrieval runs to discover.
\end{enumerate}
While IC theory is limited by its assumptions of linearity and Gaussianity and tends to struggle in low-SNR regimes where parameters remain unconstrained, it remains a powerful complementary toolkit to full Bayesian retrievals. For complex future missions like HWO, these tools provide a rapid mathematical framework to optimize instrument architectures and ensure that specific mission science objectives are met. We also emphasize that these capabilities have a further reach than just reflected light HWO observations. They could be used for any \texttt{PICASO} capability and so can be extended to  transiting exoplanets, brown dwarfs, and directly imaged planets.   We encourage the community to utilize these open-source capabilities to conduct high-dimensional trade studies across a wide range of substellar atmosphere types and multiple upcoming next-generation observatories such as HWO, LIFE, and more.

\section*{Acknowledgements}

N.E.B. and N.F.W. acknowledge partial support form the Virtual Planetary Laboratory, a member of the NASA Nexus for Exoplanet System Science (NExSS), funded via NASA Astrobiology Program grant No. 80NSSC23K1398. They also acknowledge support from the NASA Astrophysics Division.  Gemini 3.6 Flash was used for minor stylistic clarity of text and code in this paper.

\section*{Data Availability}

The spectral diagnostics package here is part of the \texttt{PICASO} v4.1 Release \citep{natasha_batalha_2026_22837176} available on Zenodo \url{https://zenodo.org/records/22837176} and GitHub \url{https://github.com/natashabatalha/picaso/releases\#release-v4.1}. Code and notebook tutorials can be seen in links throughout the manuscript.

\section*{Conflict of Interest}
Authors declare no conflict of interest.



\bibliographystyle{rasti}







\bsp	
\label{lastpage}
\end{document}